**High-pressure form of elemental boron is covalent**

N. Dubrovinskaia [1,2a)], L. Dubrovinsky [3], E. Yu. Zarechnaya [3], Y. Filinchuk [4], D. Chernyshov [4], V. Dmitriev [4], A. S. Mikhaylushkin [5], I. A. Abrikosov [5] & S. I. Simak [5]

[1]*Mineralphysik, Institut für Geowissenschaften, Universität Heidelberg, 69120 Heidelberg, Germany*

[2]*Lehrstuhl für Kristallographie, Physikalisches Institut, Universität Bayreuth, 95440 Bayreuth, Germany*

[3]*Bayerisches Geoinstitut, Universität Bayreuth, 95440 Bayreuth, Germany*

[4]*SNBL at ESRF, Boîte Postale 220, 38043 Grenoble, France*

[5]*Department of Physics, Chemistry and Biology, Linköping University, SE-581 33 Linköping, Sweden*

a)Author to whom correspondence should be addressed; electronic mail: natalia.dubrovinskaia@min.uni-heidelberg.de

**Abstract**

**Oganov et al.[1] report a discovery of an ionic high-pressure "hitherto unknown phase of boron". We show that this phase has been known since 1965, and it is a covalent material.**

Oganov et al.[1] claimed synthesis of "a hitherto unknown phase of boron". However, the first report on synthesis of the same form of boron was already published in *Science* by R.H. Wentorf[2] in 1965. The similar synthesis conditions and a perfect match of the *d*-spacings and the density of the phases reported in Refs. 1,2 allow an unambiguous conclusion that this boron phase has been known for almost 45 years. The synthesis reported by Zarechnaya et al.[3] confirms the reproducibility of Wentorf's experiments[2] and the high sample quality allowed us to solve its structure using standard methods from the X-ray diffraction data[3]. Contamination of the synthetic material could explain why Rietveld refinement failed in the work of Oganov et al.[1] and why the color of the phase was described as "dark gray"[1], while both Refs. 2 and 3 reported the characteristic deep red color in thin sections.

The central point of the paper by Oganov et al.[1] is their interpretation of the high-pressure form of boron $B_{28}$ as ionic. The authors draw this conclusion from what we regard an inappropriate analysis of the data, confusing the terms "charge transfer" (which is present to a certain degree in boron phases and elemental crystals containing symmetrically non-equivalent atoms) and "ionic" material; the outcome of this is the transformation of an ordinary result (as found in textbooks, see e.g. Fig. 19.10 of Ref. 4) into an astounding discovery. Indeed, "interstitial charge density is the characteristic feature distinguishing covalent crystals from the other two (molecular and ionic) insulating types" (Ref. 4, p. 376). Charge density redistribution (Fig. 1) due to interactions of the boron atoms constituting the $B_{28}$-phase leads to an increase of the charge density around the lines ("known in the language of chemistry as "bonds"", Ref. 4) connecting boron atoms. This is valid not only for bonds within dumbbells and icosahedra, but also for those between two icosahedra and between dumbbells and icosahedra. The electron density presented in Fig. 4a of Ref. 1 is not suggestive of ionic bonding, because it corresponds to the line connecting not neighbouring, but relatively distant atoms (~1.9 Å as compared to ~1.6 Å dumbbell-icosahedra bond). Moreover, from Fig. 1 it becomes clear that in the $B_{28}$-phase the Bader analysis ascribes the electron charge density from covalent bonds between the atoms to the atoms themselves that may result in an incorrect interpretation of the entire charge transfer picture.

Instead of performing calculations of the self-consistent electronic structure for two sublattices in B28 independently of each other (Fig. 4d in Ref. 1), which seems physically not meaningful in view of the interactions between the sublattices, we carry out a site projection of the electronic density of states. This alternative approach shows strong hybridization between the sublattices and confirms the formation of covalent bonds. Moreover, the calculated Electron Localization Function (ELF)[5] between the boron atoms has a high value (about 0.9 for its maximum in the dumbbells-icosahedra bond), a characteristic feature of strong covalent bonding. Similar high values are observed in diamond[5], in which strong covalent bonding is recognised to be a reason for diamond's extreme hardness. In a paper (Ref. 6) published prior to ref. 1, V. Solozhenko, O. Kurakevych, and A. Oganov report for $B_{28}$ a hardness of about 50 GPa, which places the material among the most superhard (hardness above 40 GPa) and thereby challenges Oganov's et al.[1] own assessment of $B_{28}$ as ionic boron boride (given that ionic superhard materials are so far unknown).

In summary, the high-pressure form of elemental boron reported by Oganov et al.[1] is neither new, nor ionic.

1. Oganov, A. R. *et al.* Ionic high-pressure form of elemental boron. *Nature*, doi:10.1038/nature07736 (*Nature* **457**, 863–867 (2009)).

2. Wentorf, R. H. Boron: another form. *Science* **147,** 49–50 (1965).

3. Zarechnaya, E. Yu. *et al.* Synthesis of an orthorhombic high pressure boron phase. *Sci. Technol. Adv. Mater.* **9**, 044209, 4pp (2008).

4. Ashcroft, N. W. & Mermin, N. D. Solid State Physics. Harcourt College Publishers, 1976.

5. Savin, A. *et al.* Electron localization in solid-state structures of the elements – the diamond structure. *Angew. Chem.Int. Ed. Eng.* **31**, 187–188 (1992).

6. Solozhenko, V. L., Kurakevych, O. O. & Oganov, A. R. On the hardness of a new boron phase, orthorhombic γ–$B_{28}$. *J. Superhard Mater.* **30**, 428–429 (2008).

**Methods**

The charge density distribution, electronic density of states and electron localization function were calculated in the framework of the same method (approximations and parameters) and simulation package as in Ref.[1] Only the energy cut-off was increased to 500 eV and integration over the Brillouin zone was performed on a fine grid of 12x12x12 special k-points.

Prismatic red crystals were synthesized at 20 GPa and 1700 K. Diffraction data were collected at SNBL on a ~6x6x25 μm crystal: space group *Pnnm*, $a$=5.0576(4), $b$=5.6245(8), $c$=6.9884(10) Å, $R_{int}$ = 0.0196, $R_1$ = 0.0387 for 37 refined parameters over 222 independent (2323 measured) reflections.

**Figure captions**

Fig. 1. The charge density distribution in $B_{28}$. (a) The difference between the calculated self-consistent charge density of $B_{28}$ and the calculated charge density resulting from the overlapping atomic densities (the "independent atomic model" in Ref. 1). The atom notations follow those used in Ref. 1. The section is drawn through the B1, which belongs to the $B_2$ dumbbell and atoms B4 and B5, belonging to the $B_{12}$ icosahedron. Formation of covalent bonds is obvious from the increase of the charge density around the lines connecting the boron atoms (most pronounced for the bond between the dumbbell atom B1 and its neighbouring atom B4 in the icosahedron). (b) Difference electron density plot around the atoms B1 and B4 extracted from experimental single-crystal X-ray diffraction data. The maximum electron density is centred in the middle of the bond, suggesting covalent bonding between the $B_2$ dumbbell and the $B_{12}$ icosahedron.

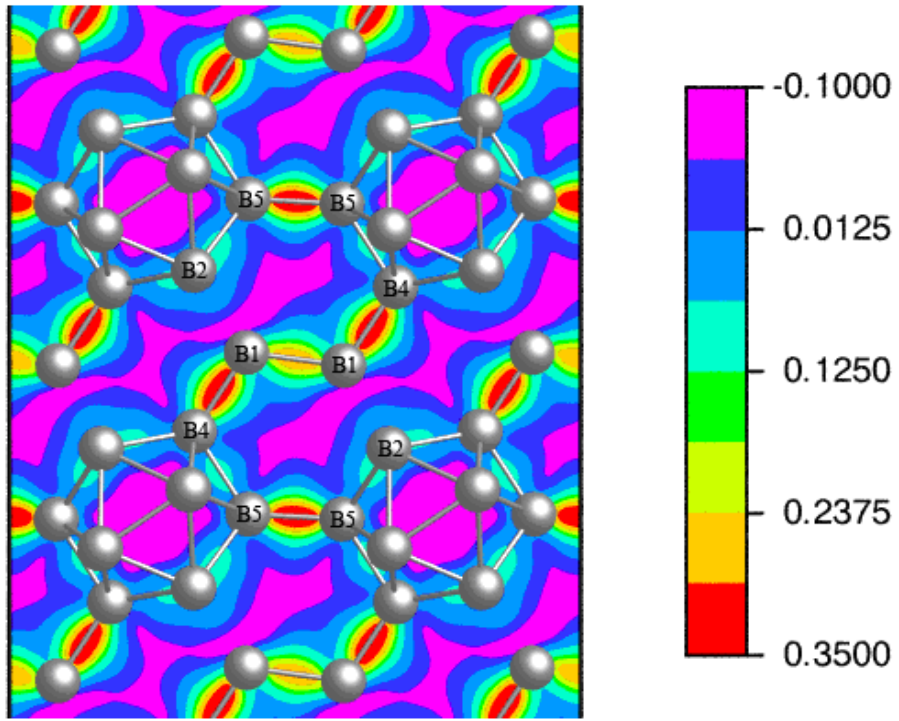

Fig. 1a.

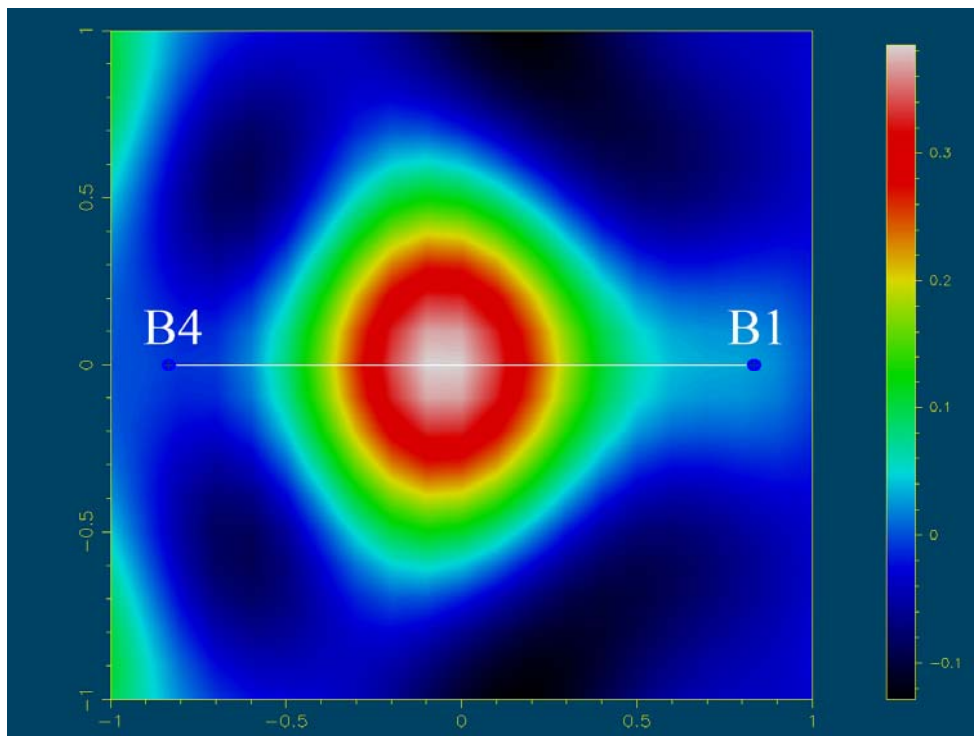

Fig. 1b.

**History of the Manuscript**

**Sent to A. Oganov prior submission** (according to *Nature* rules for Brief Communications Arising): February 11, 2009

**Submitted to *Nature*** (as Brief Communication Arising)**:**

March 1, 2009

**Reviewed**

**Rejected by *Nature* with a decision on publishing an Addendum by Oganov et al:**

May 22, 2009

**Addendum published** (doi:10.1038/nature08164; *Nature* **460**, 292 (2009))**:**

July 9, 2009